\documentclass{aa}  

\usepackage{graphicx}
\usepackage{natbib}
\bibpunct{(}{)}{;}{a}{}{,} 
\usepackage{txfonts}
\begin{document}

   \title{Gas accretion onto Jupiter mass planets in discs with laminar accretion flows}

   \author{R.P. Nelson,
          \inst{1}
          E. Lega\inst{2}
          \and
          A. Morbidelli\inst{2}}

   \institute{Astronomy Unit, Queen Mary University of London,
              Mile End Road, London, E1 4NS, U.K.\\
              \email{R.P.Nelson@qmul.ac.uk}
         \and
             Universit\'e C\^ote d'Azur, Observatoire de la Côte d'Azur, CNRS, Laboratoire Lagrange UMR7293,  Boulevard de l'Observatoire, 06304 Nice Cedex 4, France.
             \\
             \email{elena.lega@oca.eu,alessandro.morbidelli@oca.eu}
             }

   \date{Received September 5, 2022; accepted December 6, 2022}

 
  \abstract
   {Numerous studies have shown that a gap-forming Jovian mass planet embedded in a protoplanetary disc, in which a turbulent viscosity operates, can accrete gas efficiently through the gap, and for typical parameters it doubles its mass in $\sim 0.1$~Myr. The planet also migrates inwards on a timescale that is closely related to the local viscous evolution timescale, which is also typically 0.1~Myr. These timescales are short compared to protoplanetary disc lifetimes, and raise questions about the origins of the cold gas giant exoplanets that have been discovered in abundance. It is understood that protoplanetary discs are unlikely to be globally turbulent, and instead they may launch magnetised winds such that accretion towards the star occurs in laminar accretion flows located in narrow layers near the surfaces of the disc.}
   {The aim of this study is to examine the rate at which gas accretes onto Jovian mass planets that are embedded in layered protoplanetary discs, and to compare the results with those obtained for viscous models. }
   {We use 3D hydrodynamical simulations of planets embedded in protoplanetary discs, in which a constant radial mass flux towards the star of ${\dot m} = 10^{-8}$~M$_{\odot}$~yr$^{-1}$ is sustained. We consider a classical viscous $\alpha$ disc model, and also models in which an external torque is applied in narrow surface layers to mimic the effects of a magnetised wind. The accreting layers have a variety of depths, as parameterised by their column densities $\Sigma_{\rm A}$, and we consider values of $\Sigma_{\rm A}$ in the range 0.1 to 10~g~cm$^{-2}$.}
   {The viscous disc model gives results in agreement with previous studies. In accord with our recent work that examines the migration of Jovian mass planets in layered models, we find the accretion rate onto the planet in the layered models crucially depends on the ability of the planet to block the wind-induced mass flow towards the star. For $\Sigma_{\rm A}=10$~g~cm$^{-2}$, the planet torque can block the mass flow in disc, accretion onto the planet is slow, and a mass doubling time of 10~Myr is obtained. For $\Sigma_{\rm A}=0.1$~g~cm$^{-2}$, the flow is not blocked, accretion is fast, and the mass doubling time is 0.2~Myr}
   {Our results show that although the radial mass flow through the layered disc models is always $10^{-8}$~M$_{\odot}$~yr$^{-1}$, adopting different values of $\Sigma_{\rm A}$ leads to very different gas accretion rates onto embedded gas giant planets.}
   \titlerunning{Gas accretion onto planets in discs with laminar accretion flows.}
  \authorrunning{R.P. Nelson et al.}

   \maketitle
%

\section{Introduction}
Observational surveys have demonstrated that the exoplanet population is dominated by super-Earths and mini-Neptunes, while giant planets make up a sub-population that accounts for only $\sim 10\%$ of the overall planet number \citep{2013ApJS..204...24B, 2011arXiv1109.2497M}. Efforts have been made to construct an unbiased sample of giant planets based on radial velocity discoveries, and these suggest that the distribution of orbital periods has two maxima, separated by a valley that sits in the range $10 \le P \le 100$~days \citep{2008PASP..120..531C,2016A&A...587A..64S}. A significant majority of giant planets orbit with periods longer than 100 days, and we refer to these as cold Jupiters. The masses of the giant planets discovered by radial velocities show a fairly flat distribution between $0.5 \le M_{\rm p} \sin{i} \le 2$~M$_{\rm Jup}$, which then declines down to $\sim 4$~M$_{\rm Jup}$ before flattening off at larger masses.\footnote{These statements are based on data obtained from https://exoplanetarchive.ipac.caltech.edu/ downloaded on 26 May 2022.} Currently, we do not have a good theoretical understanding of the mass and orbital period distributions of giant exoplanets.

The giant planet sub-population can be considered to be distinct from the broader population of exoplanets for two physically motivated reasons. First, the core-accretion model shows that Jovian mass planets accrete the majority of their gas during a runaway phase of growth, whereas mini-Neptunes and Neptune-like planets accrete their gas envelopes quasi-statically \citep{1996Icar..124...62P}. This means that the rate at which gas is supplied to the planet by the protoplanetary disc is likely to be an important factor in determining the masses of giant planets. Second, gas giant planets are believed to open deep gaps in protoplanetary discs \citep{1979MNRAS.186..799L,1986ApJ...307..395L,1980ApJ...241..425G}, and this can strongly influence both the gas accretion and the orbital migration rates of these bodies \citep{1999ApJ...514..344B,1999MNRAS.303..696K}.

Giant planets are believed to start forming close to the snow line, where the condensation of water enhances the density of solid material and enables the rapid formation of the first planetesimals \citep{2016A&A...596L...3I,2017A&A...602A..21S,2017A&A...608A..92D}. The location of the snow line in young discs is expected to be at 5--7~au \citep{2021A&A...648A.101L, 2020ApJ...901..166V}, indicating that the cold Jupiter population migrated only a few astronomical units in the protoplanetary disc, which is at odds with the predictions of most studies of disc-driven giant planet migration. Giant planets are expected to form a gap in the disc and undergo type II migration. In the traditional picture, this mode of migration occurs with the planet migrating within the gap at the viscous evolution rate of the disc. Accretion rates onto young stars have typical values ${\dot m} \sim 10^{-8}$~M$_{\odot}$~yr$^{-1}$ \citep{1998ApJ...495..385H}, which requires the viscosity parameter to have values $\alpha \gtrsim 10^{-3}$, giving rise to migration timescales $\sim 0.1$~Myr \citep{2000MNRAS.318...18N}. Given that discs have typical lifetimes of a few million years. \citep{2001ApJ...553L.153H} this makes it difficult to explain the cold Jupiter population, and indeed models show that a giant planet needs to start undergoing type II migration at $\sim 15-20$~au from the star in order to halt at a distance of a few astronomical units after a few million years of migration \citep{2014MNRAS.445..479C,2015A&A...582A.112B}. Although the classical view of type II migration has been revisited recently \citep{2015A&A...574A..52D, 2018A&A...617A..98R}, such that the migration speed is not exactly equal to the unperturbed drift speed of the disc, it is still proportional to the viscosity when $\alpha \ge 10^{-4}$, and a significant reduction in $\alpha$ is required to reduce the distance over which type II migration occurs over a disc's lifetime \citep{2021A&A...646A.166L}.

Despite the presence of a gap, numerous studies of gas accretion onto Jovian mass planets show that a  mass flux through the gap of ${\dot m} \sim 10$~M$_{\rm Jup}$~Myr$^{-1}$ is sustained for $\alpha \ge 10^{-3}$, essentially corresponding to the unperturbed accretion flow through the disc \citep{1999ApJ...514..344B,1999MNRAS.303..696K,1999ApJ...526.1001L}. Hence, if the planet can accrete a large fraction of the gas that is supplied to it, then it is difficult to explain the giant planet mass distribution, since the mass doubling time of a Jupiter-mass body is only $\sim 10^5$~yr. Recent studies have shown that thermodynamic effects in the planet's Hill sphere may be important for slowing down gas accretion under some circumstances \citep[e.g.][]{2016MNRAS.460.2853S, 2021A&A...646L..11M, 2022arXiv220211422M}; however, the radiation-hydrodynamic simulations of \cite{2019A&A...630A..82L} indicate that Jovian mass planets can accrete at a rate ${\rm M}_{\rm p} \ge 10$~M$_{\rm Jup}$~Myr$^{-1}$. It is possible that other effects such as magnetic fields may influence the accretion rate, but the non-ideal magnetohydrodynamical (MHD) study of gas accretion onto giant planets by \cite{2013ApJ...779...59G} showed accretion rates in line with the values quoted above. Hence, within the context of viscous disc models, it is difficult to explain either the mass or the period distribution of the extrasolar giant planets.

The above discussion applies to planets embedded in viscous discs, where it was originally believed that the viscosity in discs might arise from the magneto-rotational instability \citep[MRI,][]{1991ApJ...376..214B}. The very low ionisation fraction in the main body of a protoplanet disc, however, quenches the MRI \citep{1996ApJ...457..355G} and instead a combination of global magnetic fields, non-ideal MHD effects and the ionisation of disc surface layers can launch magnetically driven disc winds that also drive a laminar accretion flow through the surface layers of the disc \citep{2013ApJ...769...76B,2015ApJ...801...84G,2017A&A...600A..75B}. Hence, planet formation may occur in discs with very low levels of turbulence and where accretion towards the star occurs in narrow surface layers that sustain rapid radial gas flows.

Motivated by this change in our understanding of the internal dynamics of protoplanetary discs, we have recently presented simulations of giant planets migrating in very low viscosity discs. In \cite{2021A&A...646A.166L}, we examined migration in discs without wind-driven accretion flows, and showed that initially a giant planet migrates inwards because a vortex forms at the outer gap edge, but once this vortex dissipates the `vortex-driven migration' ceases and migration essentially halts. In \cite{2022A&A...658A..32L}, we examined the migration of giant planets in discs with wind-driven laminar accretion flows, and we showed that the behaviour crucially depends on whether or not the torque from the planet is strong enough to block the accretion flow. When the flow is fast and occurs relatively unimpeded by the planet, we found that migration is slow and occurs at a speed ${\dot r_{\rm p}} \sim 3$~au~Myr$^{-1}$. When flow is blocked by the planet then migration can be fast, occurring at a speed ${\dot r_{\rm p}} \sim 15$~au~Myr$^{-1}$. Hence, for appropriate disc parameters, giant planets undergoing slow migration can explain the cold Jupiter population. 

In this study we examine the rate of gas accretion onto giant planets that are held on fixed circular orbits, and which are embedded in layered disc models with laminar accretion flows that are very similar to those considered in \cite{2022A&A...658A..32L}. Our aim is to determine the conditions under which the gas flow towards the planet can be significantly decreased by the planet's tidal torques, such that the planet regulates the rate at which it can accrete gas, and to understand how the joint mass and orbital evolution of a gas giant planet proceeds in layered disc models.

The paper is organised as follows. In Sect.~\ref{sec:Methods} we present the basic equations, physical model and numerical methods. In Sect.~\ref{sec:Theory} we discuss our theoretical expectations, and in Sect.~\ref{sec:Results} we present the simulation results. In Sect.~\ref{sec:Discussion} we discuss the results and their implications for the formation and early evolution of giant plates, and in Sect.\ref{sec:Conclusions} we draw our conclusions.

\begin{figure*}
\includegraphics[width=0.33\textwidth]{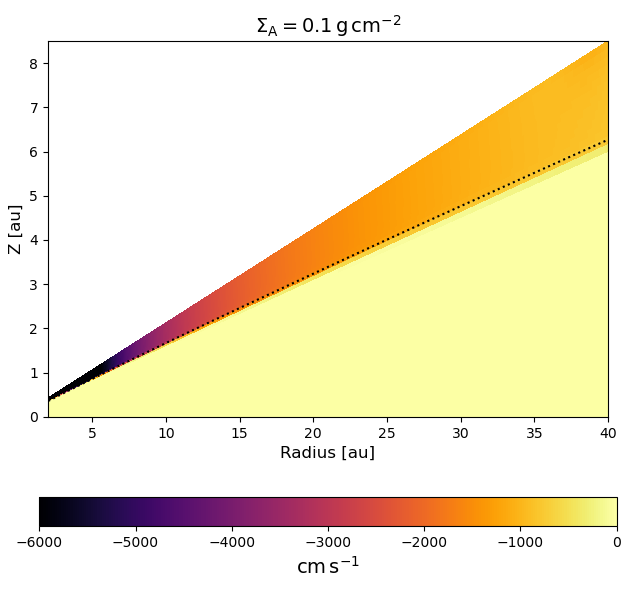}
\includegraphics[width=0.33\textwidth]{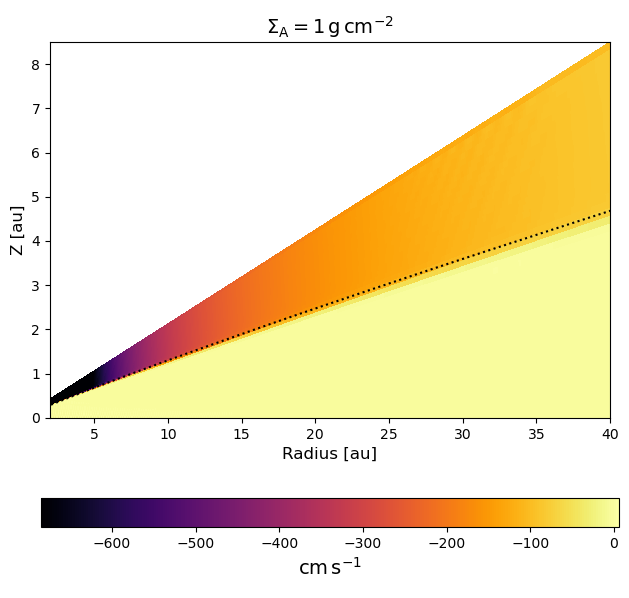}
\includegraphics[width=0.33\textwidth]{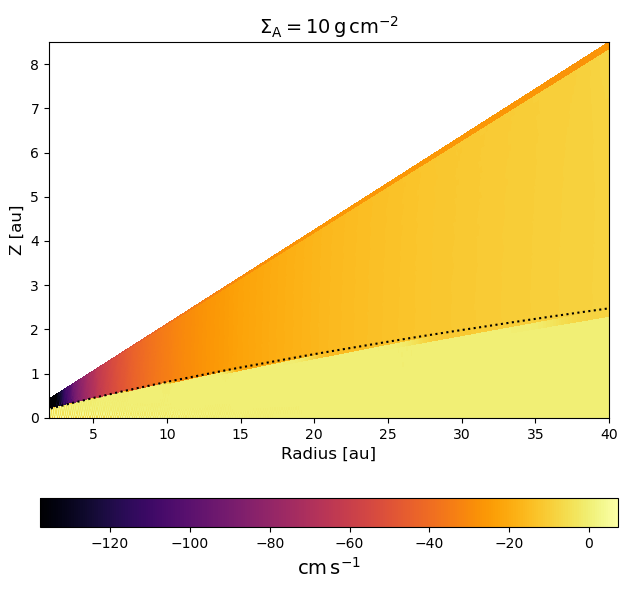}
\caption{
Contour plots showing the radial velocity distributions in the various layered disc models prior to insertion of the planets. The black dotted line indicates the location where the column density measured from the disc surface corresponds to $\Sigma_{\rm A}$. We note that only the upper hemisphere of the disc is simulated.
}
\label{fig:VrContours}
\end{figure*}

\section{Basic equations and numerical methods}
\label{sec:Methods}
\subsection{Equations of motion}
\label{sec:equations}
We use a variety of coordinate systems in this paper. Cartesian coordinates are denoted by ($x$, $y$, $z$), spherical polar coordinates by ($r$, $\theta$, $\phi$) and cylindrical polar coordinates by ($R$, $\phi$, $z$). We solve the continuity, momentum and internal energy equations 
\begin{eqnarray}
\frac{\partial \rho}{\partial t} + \nabla \cdot (\rho {\bf v}) & = & 0 \nonumber \\
\frac{\partial {\bf v}}{\partial t} + ({\bf v} \cdot \nabla) {\bf v} & = & - \frac{1}{\rho}\nabla P - \nabla \Phi + {\bf S}_{\rm visc} + f_{\rm wind} {\bf \hat \phi} \nonumber \\
\frac{\partial e}{\partial t} + \nabla \cdot (e {\bf v}) & = & -{P} \left(\nabla \cdot {\bf v}\right) - \rho c_{\rm V} \frac{\left(T-T_{\rm i}\right)}{\tau_{\rm cool}}, 
\label{eqn:cont+mom+en}
\end{eqnarray}  
where ${\bf v}$ is the velocity, $\rho$ is the density, $P$ is the pressure, $e$ is the internal energy per unit volume and $T$ is the temperature ($T_{\rm i}$ being its initial value and $\tau_{\rm cool}$ being the cooling timescale on which temperature fluctuations relax towards $T_{\rm i}$). The viscous force per unit mass is denoted by ${\bf S}_{\rm visc}$, and $f_{\rm wind} {\bf \hat \phi}$ is an azimuthal acceleration applied to the disc designed to mimic the back reaction on the disc when a magnetised wind is centrifugally launched from the disc surface ($\bf \hat \phi$ being the unit vector in the $\phi$ direction). In this work we adopt an ideal gas equation of state, $P=(\gamma-1) e$ with $\gamma=7/5$, and we set $\tau_{\rm cool}=1$, in units of the local orbit period, as this timescale is long enough for the vertical shear instability (VSI) to be suppressed \citep{2013MNRAS.435.2610N}. Recent radiation-hydrodynamic simulations that adopt realistic opacities indicate that the VSI does not operate in the inner regions ($\lesssim 10$ au) of protoplanetary discs \citep{2017ApJ...850..131F}, hence our choice for $\tau_{\rm cool}$ is justified as this is the region of interest in this study. 

The gravitational potential, $\Phi$, arises from the central star and the planet (which is maintained on a fixed circular orbit), and is given by
\begin{equation}
\Phi({\bf r}) = -\frac{G M_*}{r} + \Phi_{\rm p} +\Phi_{\rm ind},
\label{eqn:Phi}
\end{equation}
where $\Phi_{\rm ind}$ is an indirect term that arises because we work in a non-inertial frame centred on the star, and which includes contributions from the planet and the disc. $\Phi_{\rm p}$ is the potential due to the planet, and we use the prescription of \cite{2009A&A...506..971K} in our implementation of this term:
\begin{equation}
\Phi_{\rm p} =
\begin{cases}
- \frac{G M_{\rm p}}{d} & d > \epsilon \\ \\
- \frac{G M_{\rm p}}{d} f\left(\frac{d}{\epsilon}\right) & d \le \epsilon
\end{cases}
\label{Phi_p}
\end{equation}
where $d=|{\bf r} - {\bf r}_{\rm p}|$ and $f(x)=x^4 -2x^3+2x$. We adopt a softening length $\epsilon=0.4 R_{\rm H}$, where $R_{\rm H}=r_{\rm p} \left(M_{\rm p}/3 M_*\right)^{1/3}$ is the Hill radius.

\subsection{Numerical methods}
\label{sec:numerics}
The above set of equations are solved using the 3-dimensional hydrodynamical code NIRVANA \citep{1997CoPhC.101...54Z}  in spherical polar coordinates ($r$, $\theta$, $\phi$). This code has been used extensively in the study of protoplanetary discs, with and without embedded planets \citep[e.g.][]{2004MNRAS.350..849N, 2006A&A...457..343F,2013MNRAS.435.2610N}. NIRVANA uses an algorithm similar to the ZEUS code \citep{1992ApJS...80..753S}. The equations are divided into source and transport terms, and operator splitting is used to update the state variables according to the different terms in this formalism. The \cite{1977JCoPh..23..263V} upwind, monotonic advection scheme is employed to update the transport terms. We have also utilised the FARGO3D code \citep{2016ApJS..223...11B} to conduct simulations to compare with and validate the NIRVANA runs, and have obtained very similar results using the two codes. A comparison of the two codes is presented in Appendix~\ref{sec:Appendix}.

As discussed below, we employ disc models that have initial temperature distributions that give rise to constant aspect ratios, $h \equiv H/r$, where $H$ is the vertical scale height. The computational domain in the meridional direction extends from the disc surface, located at 4 scale heights above the midplane, down to the disc midplane at $\theta=\pi/2$. A symmetry boundary condition is employed at the midplane, and standard outflow conditions are applied at the disc surface. The azimuthal domain lies in the range $0 \le \phi \le 2 \pi$. The radial domain covers the interval $1.04 \le r \le 46.8$, where the unit of length is the astronomical unit (au). Further details about the system of units employed in this paper are given below. The planet is inserted at a radius of $r_{\rm p}=5.2$~au in all simulations, and remains on a fixed circular orbit. We employ wave damping boundary conditions near both the inner and outer radial boundaries, which are implemented by damping the density, radial and vertical velocities towards their initial values on a timescale of $0.1$ local orbital periods \citep[for more details see][]{2006MNRAS.370..529D}. In addition, a linear bulk viscosity is employed near the outer boundary to provide additional damping. The implementation of this term is the same as in the ZEUS code \citep{1992ApJS...80..753S}. In our runs the linear viscosity coefficient, $Q_{\rm lin}$, ramps up linearly from 0 to 1 between the inner edge of the damping zone and the outer edge of the computational domain. Wave-damping is employed in the region $r \le 1.612$~au, and both wave-damping and linear viscosity are applied in the region $r \ge 35.672$~au. The numerical resolution we adopt in all simulations is ($N_r$, $N_{\theta}$, $N_{\phi}$)=(576, 48, 384), giving a uniform radial grid spacing $\Delta r\simeq 0.08$.

The simulations are conducted in a reference frame that corotates with the planet. We initialise the disc models according to the physical parameters described below, and allow a disc to relax for more than 200 orbits at the location of the planet before the latter is inserted, in order to allow a steady mass flow to be established. Upon insertion, the planet-star mass ratio $q_{\rm p}=M_{\rm p}/M_*=6 \times 10^{-5}$. Accretion of gas onto the planet occurs according to
\begin{equation}
\frac{d M_{\rm p}}{dt} = \sum_i K \, \rho_i \, \Delta V_i,
\label{eqn:dmpdt}
\end{equation}
where $K$ is a constant (with $0 < Kdt < 1 $) and $\rho_i \, \Delta V_i$ is the mass contained in a grid cell, and the summation occurs over all cells that are located within a distance equal to half the planet's Hill sphere radius. The accreted mass is removed from the disc and is added to the planet's mass until $q_{\rm p}=M_{\rm p}/M_*=10^{-3}$ (corresponding approximately to the Jovian mass), after which time the planet's mass is held fixed while the rate of gas accretion continues to be monitored. This procedure for introducing and growing the planet is adopted to ensure the disc is not shocked by the sudden introduction of a giant planet, to conserve mass during the planetary growth and gap formation phase, and also as a means of crudely mimicking the runaway gas accretion that is expected to occur onto a planet with $q_{\rm p}=6 \times 10^{-5}$ as it grows to become a gas giant. We initiate the simulations using the value $K=6$ in eq.~\ref{eqn:dmpdt}, and at late times this is increased to $K=30$. This was done because in earlier runs, adopting $K=6$ did not lead to a clear separation of the long-term accretion rates onto the planets for different disc models, and in particular did not reflect the rate at which mass is supplied to the planet by the background disc. This problem is ameliorated by increasing the value of $K$, and we found that increasing the value to $K=30$ led to a clear separation of the accretion rates experienced by the planets in the different runs, and to values that are in line with expectations given the rate of mass supply to the planet. Hence, we did not experiment with further values of $K$ to examine the effects of varying this somewhat arbitrary parameter. Understanding what value of $K$ is `realistic' remains an area of active research \citep[see][for example]{2022arXiv220309595P}, and a discussion about related issues is provided in Sect.~\ref{sec:GasSupply}.

We note that our simulations do not resolve the structure of the gas flow within the planet's Hill sphere because $R_{\rm H} /\Delta r \simeq 4.6$ when $q_{\rm p}=10^{-3}$. Hence, the simulations are designed to test the ability of the planet's tidal torques to retard the large scale accretion flow of the gas into the gap and then into the Hill sphere, and do not address the question of how the gas then accretes onto the actual planet (see Sect.~\ref{sec:GasSupply} for a discussion about  gas supply to the Hill sphere and the planet's accretion rate), which at this stage of the evolution is expected to have contracted to a size that is significantly smaller than the Hill sphere radius \citep[e.g.][]{2005A&A...433..247P}.  
We further note that because we only simulate the upper hemisphere of the disc models, when quoting accreted masses, accretion rates onto the planet, or when adding mass to the planet due to accretion, we multiply the accreted mass by a factor of two to account for the mass that would be accreted from the missing hemisphere.

\begin{figure*}
\includegraphics[width=0.33\textwidth]{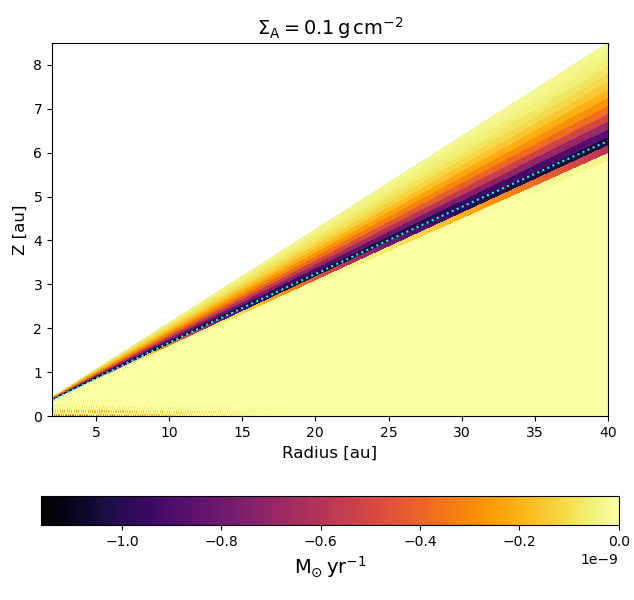}
\includegraphics[width=0.33\textwidth]{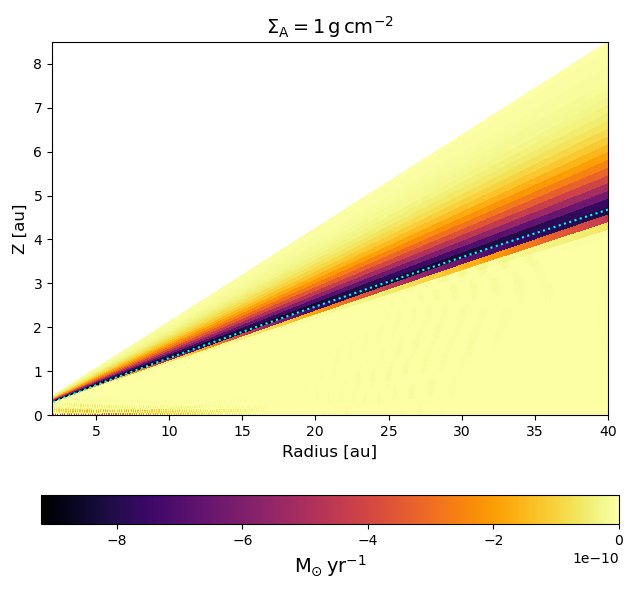}
\includegraphics[width=0.33\textwidth]{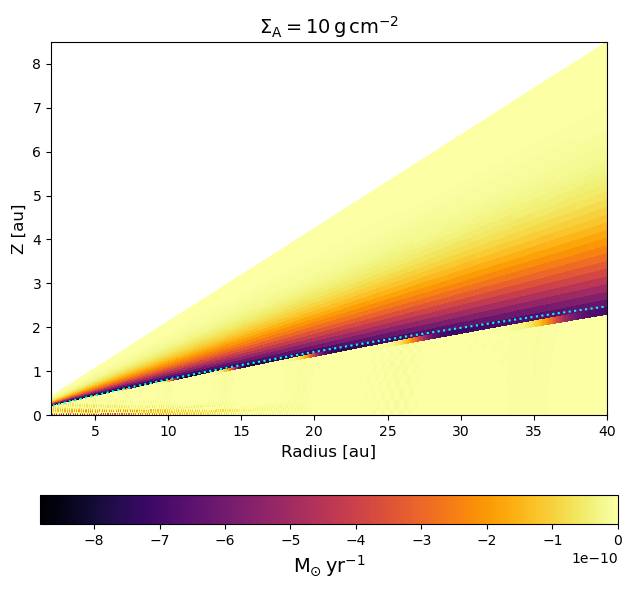}
\caption{
Contour plots showing time-averaged and azimuthally-integrated radial mass flux distributions in the various layered disc models. The cyan dotted line indicates the location where the column density measured from the disc surface corresponds to $\Sigma_{\rm A}$. Note that only the upper hemisphere of the disc is simulated.
}
\label{fig:MfluxContours}
\end{figure*}

\subsection{Disc models with spatially constant accretion rates}
We adopt disc models for which the surface density, defined by
\begin{equation} 
\Sigma = \int_{-\infty}^{\infty} \rho \, dz, 
\label{eqn:Sigma}
\end{equation}
is initially axisymmetric. The midplane density is given by a power law in cylindrical radius
\begin{equation}
\rho(R,z=0)=\rho_0\left(\frac{R}{R_0}\right)^p,
\label{eqn:rho}
\end{equation}
and the temperature is given by
\begin{equation}
T(R,z) = T_0 \left( \frac{R}{R_0} \right)^q.
\label{eqn:T}
\end{equation}
Expressions for the equilibrium densities and azimuthal velocities for these power-law disc models are given in \cite{2013MNRAS.435.2610N}.
The isothermal sound speed $c_{\rm s} = \sqrt{{\cal R} T /\mu}$, and the pressure scale height of the disc $H=c_{\rm s}/\Omega_{\rm K}$, where $\Omega_{\rm K}$ is the keplerian angular velocity. We adopt the values $p=-3/2$ and $q=-1$ for the power-law indices in equations~(\ref{eqn:rho}) and (\ref{eqn:T}), and a value of $T_0$ that gives an aspect ratio $h=0.05$ throughout the disc. The value of $\rho_0$ is chosen so that $\Sigma= \sqrt{2 \pi} \rho_{z=0} H = 220$ g cm$^{-2}$ at $r=5.2$~au (note that this value includes both the top and bottom hemispheres of the disc; the column density from the surface to the midplane is 110 g cm$^{-2}$).  We further note the surface density at 5 au in the minimum mass solar nebula has a value $\sim 150$ g cm$^{-2}$ \citep{1981PThPS..70...35H}.

Test simulations have shown that when completely inviscid discs are considered, convergence between simulations performed at different resolutions can be unsatisfactory. Adding a small viscosity significantly improves the situation \citep{2022A&A...658A..32L}, so we include an alpha viscosity with $\alpha=10^{-5}$ in our layered disc models and in a calibration run which does not contain an accreting layer (the No Wind simulation). 

\subsubsection{Layered accretion due to an external torque}
To obtain an expression for the azimuthal acceleration, $f_{\rm wind}$ (first introduced in equation~\ref{eqn:cont+mom+en}), that drives a constant mass flux in a laminar disc, we first note that the mass flux at each cylindrical radius, $R$, is given by
\begin{equation}
{\dot m} =  2 \pi R \Sigma_{\rm A} | v_R |,
\label{eqn:mdot}
\end{equation}
where $\Sigma_{\rm A}$ is the column density associated only with the actively accreting layer of the disc. When considering layered accretion, $\Sigma_{\rm A} < \Sigma$ holds except when the whole vertical column of the disc is active and accreting.
The specific angular momentum in a keplerian disc  $j=\sqrt{GM_* R}$, and differentiating with respect to time gives
\begin{equation}
v_R = 2 \sqrt{\frac{R}{G M_*}} \frac{\partial j}{\partial t}.
\label{eqn:v_R}  
\end{equation}
Combining equations~(\ref{eqn:mdot}) and (\ref{eqn:v_R}) gives
\begin{equation}
f'_{\rm wind} {\bf \hat \phi} = - \sqrt{\frac{G M_*}{R^5}} \frac{{\dot m}}{4 \pi \Sigma_{\rm A}},
\label{eqn:f_wind}
\end{equation}
where $\partial j/\partial t = R\, f'_{\rm wind} {\bf \hat \phi}$. Equation~(\ref{eqn:f_wind}) can be used to set up a disc with a chosen spatially and temporally constant radial mass flux, where the mass flux is confined to the surface layers where $\Sigma(z) \le \Sigma_{\rm A}$, where $\Sigma(z)$ is the column density measured from some height in the disc, $z$, up to the disc surface
\begin{equation}
    \Sigma(z)=\int_{z}^{\infty} \rho dz.
\end{equation}
We refer to the torque that arises from $f'_{\rm wind}$ as the `wind-driven torque', in recognition of the fact that this is supposed to mimic the torque that arises from the launching of a magnetised wind.

Our numerical implementation of equation~(\ref{eqn:f_wind}) involves calculating the column density, $\Sigma$, at each position in the disc every 100 time steps, and then applying the following acceleration, which employs the Fermi function to transition the applied acceleration across the boundary between the accreting and non-accreting layers of the disc:
\begin{equation}
f_{\rm wind} = f'_{\rm wind} \cdot \left( \frac{1}{\exp{\left(\frac{\Sigma(z) - \Sigma_{\rm A}}{\Sigma_{\rm T}}\right) + 1}} \right),
\label{eqn:fermi_function}
\end{equation}
where $\Sigma_{\rm T} \ll \Sigma_{\rm A}$ sets the width of the transition between the active and dead layers. A further consideration is how mass is to be supplied at the disc outer radius so that it can maintain a constant mass flux over time. The use of a buffer zone at the outer disc edge, in which the density is continuously relaxed towards its initial value on sub-orbital timescales, ensures that our disc models maintain constant accretion rates through the main body of the disc at all times. As a final comment concerning the implementation of the wind-driven torque, we note that the effects of mass loss from the disc through the magnetocentrifugally driven wind are not included. Generally, the mass loss rate through the wind is significantly smaller than the mass flow rate in the accretion flow, and the run times we consider are too short for the mass loss in the wind to affect the global disc structure.

Equation~\ref{eqn:f_wind} shows that for a given radial mass flux through the disc, ${\dot m}$, the applied acceleration is inversely proportional to the column density of the active layer, $\Sigma_{\rm A}$. In this work we consider a single value of ${\dot m} = 10^{-8}$~M$_{\odot}$~yr$^{-1}$ (each hemisphere of the disc provides half of this value) and values of $\Sigma_{\rm A}=0.1$, 1 and 10~g~cm$^{-2}$. We adopt $\Sigma_{\rm T}=0.1$ for the $\Sigma_{\rm A}=1$ and 10~g~cm$^{-2}$ cases, and $\Sigma_{\rm T}=0.01$ for the $\Sigma_{\rm A}=0.1$~g~cm$^{-2}$ case.
The steady state velocity profiles that arise from the application of eqns~(\ref{eqn:f_wind}) and (\ref{eqn:fermi_function}) are illustrated by fig.~\ref{fig:VrContours}, and as expected the radial velocity scales inversely with $\Sigma_{\rm A}$. For reference, we note that at $r=5.2$~au the isothermal sound speed $c_{\rm s}\simeq0.65$~km~s$^{-1}$ and the Mach number of the radial accretion flow has values ${\cal M} \simeq 10^{-1}$, $10^{-2}$ and $10^{-3}$ for $\Sigma_{\rm A}=0.1$, $1$ and $10$~g~cm$^{-2}$, respectively. 

\begin{figure}
\includegraphics[width=9cm]{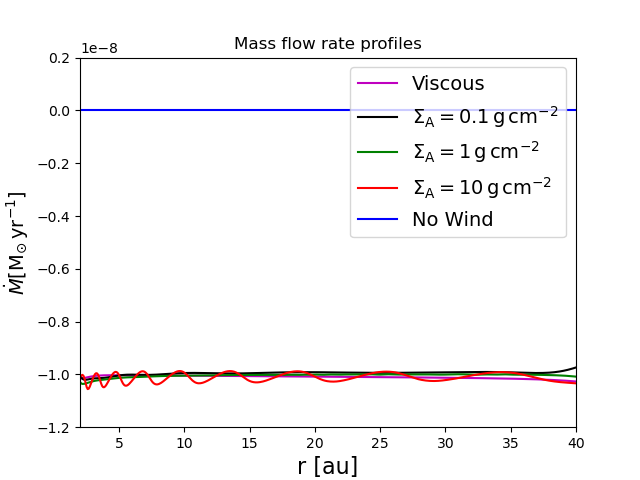}
\caption{
Radial profiles of the gas accretion rates in the viscous and externally torqued disc models.
}
\label{fig:MdotProfiles}
\end{figure}

Figure~\ref{fig:MfluxContours} shows the azimuthally integrated mass fluxes corresponding to the velocities shown in fig.~\ref{fig:VrContours}, and vertically integrated radial profiles of the mass fluxes for the layered and viscous models (described below) are shown in Figure~\ref{fig:MdotProfiles}, where it can be seen that both hemispheres of the disc combined produce a radial mass flux of  $10^{-8}$ M$_{\odot}$ yr$^{-1}$. Hence, mass will be supplied to embedded planets by the large scale background accretion flow in the unperturbed discs at a rate of $10^{-8}$ M$_{\odot}$ yr$^{-1}$, equivalent to $10$ M$_{\rm Jup}$ Myr$^{-1}$. The mass doubling time for a Jupiter mass planet being fed by such a flow would then be $10^5$ yr, similar to the values obtained by numerous previous simulations of planets embedded in viscous discs \citep[e.g.][]{1999ApJ...514..344B,1999MNRAS.303..696K,1999ApJ...526.1001L}. We note that the oscillations in the mass flux for the $\Sigma_{\rm A}=10$ case in Figs.~\ref{fig:MfluxContours} and \ref{fig:MdotProfiles} arise because the width adopted for the transition between the active and dead zones is only marginally resolved in this run. Although this is unlikely to adversely impact the results of the simulations presented here, we have confirmed that widening the transition removes these oscillations.

\subsubsection{Viscous models}
In addition to the externally torqued laminar models discussed in the previous section, for comparison purposes we also compute a viscous disc model that has the same radial mass flux. From standard viscous thin disc theory, we have the following expression for the torque per unit mass acting at each radius in the disc
\begin{equation}
v_R \frac{\partial}{\partial R} \left(R^2 \Omega \right) =  \frac{1}{R \Sigma} \frac{\partial }{\partial R} \left( R^3 \nu \Sigma \frac{d \Omega}{dR} \right),
\label{eqn:visc-theory}
\end{equation}
where $\nu$ is the kinematic viscosity. Given ${\dot m} = -2 \pi R \Sigma v_R$, we can write
\begin{equation}
-\frac{{\dot m}}{2\pi} =   \left(\frac{\partial}{\partial R} \left(R^2 \Omega\right) \right)^{-1} \frac{\partial}{\partial R} \left( R^3 \nu \Sigma \frac{d\Omega}{dR} \right).
\label{eqn:mdot-2pi}
\end{equation}
If we adopt power-laws for the surface density and kinematic viscosity profiles, $\Sigma(R) = \Sigma_0 R^{\beta}$ and $\nu(R) = \nu_0 R^{\delta}$, and consider a keplerian disc with $\Omega = \sqrt{G M_*/R^3}$, then we obtain the result that ${\dot m}$ is independent of radius when $\delta=-\beta$. Hence, for the disc models with $\Sigma \propto R^{-1/2}$ considered in this paper, we require $\delta=1/2$. Equation~(\ref{eqn:mdot-2pi}) also leads to the expression
\begin{equation}
{\dot m} =  3 \pi \nu_0 \Sigma_0,
\label{eqn:3pi-nu-sigma}
\end{equation}
such that the kinematic viscosity needed to produce the required mass accretion rate can be determined once the surface density has been specified. The steady radial mass flux obtained from the viscous disc model we compute is shown in figure~\ref{fig:MdotProfiles}, where the magenta line is seen to sit under the lines for the other models at a value of ${\dot m}=10^{-8}$~M$_{\odot}$~yr$^{-1}$, as required.

Finally, we note in passing that the requirement for $\delta=1/2$ in a disc with $\Sigma=\Sigma_0 R^{-1/2}$ is satisfied by the $\alpha$ model for the kinematic viscosity, $\nu = \alpha H^2 \Omega$, when $H/r$ is constant, as is the case in our models. 

\subsection{Summary of the runs}
We present the results of five simulations in this paper: a No Wind simulation that does not have a laminar accretion flow and where a very small viscosity ($\alpha=10^{-5}$) is employed; a Viscous run in which the radial mass flux ${\dot m}=10^{-8}$~M$_{\odot}$~yr$^{-1}$ at all radii; three layered models that contain laminar accretion flows with ${\dot m}=10^{-8}$~M$_{\odot}$~yr$^{-1}$ at all radii, and with $\Sigma_{\rm A}=0.1$, 1 and 10~g~cm$^{-2}$.
\subsection{Units}
When discussing simulation results the unit of length is 1~au, the unit of time is years and the unit of mass is the solar mass or Jupiter's mass (depending on the context, and where $M_{\rm Jup}=10^{-3} M_{\odot}$). Surface densities, volume densities and velocities are generally quoted in cgs units.

\section{Theoretical expectations}
\label{sec:Theory}
The long term steady gas accretion rate onto a planet embedded in the disc models should be determined by the balance between the tidal torque exerted on the disc by the planet, and the viscous or wind-driven torque acting to drive the radial mass flux through the disc. The viscous torque per unit mass is given by eqn.~(\ref{eqn:visc-theory}) and combining this with $\Lambda_{\rm p}$, the torque per unit mass due to the planet, leads to the following expression for the radial mass flux in a viscous disc containing an embedded planet
\begin{equation}
-\frac{{\dot m}}{2\pi} =   \left(\frac{\partial}{\partial R} \left(R^2 \Omega\right) \right)^{-1} \left[ \frac{\partial}{\partial R} \left( R^3 \nu \Sigma \frac{d\Omega}{dR} \right) + R \Sigma \Lambda_{\rm p} \right].
\label{eqn:mdot-2pi+planet}
\end{equation}
An embedded planet can open a gap in the disc when tidal torques locally exceed the viscous torques, and eqn.~(\ref{eqn:mdot-2pi+planet}) shows that the viscous torque readjusts to the changing disc structure such that in principle viscous and planet torques can cancel each other, resulting in ${\dot m}=0$, with gas accretion onto the planet being switched off. The torque from the planet is highly localised, however, so the mass flux far from the planet is relatively unaffected and hence the gap structure must evolve on longer timescales as gas flows towards the planet, such that torque balance cannot be maintained. For a planet on a fixed circular orbit, a steady state then corresponds to mass accreting through the gap towards the planet at the rate supplied through the disc, part of which will be accreted by the planet. Indeed, numerous hydrodynamical simulations of planets embedded in viscous discs have shown that accretion through gaps is maintained at essentially the rate at which gas flows through the unperturbed disc \citep{1999ApJ...514..344B,1999MNRAS.303..696K,1999ApJ...526.1001L}. Hence, in our viscous model we expect to see steady accretion onto the planet at a rate of $\sim 10^{-8}$~M$_{\odot}$~yr$^{-1}$ (corresponding to 10~M$_{\rm Jup}$ Myr$^{-1}$).
\begin{figure}
\includegraphics[width=\columnwidth]{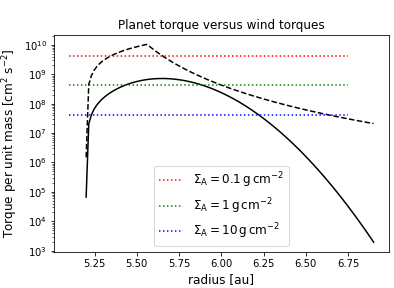}
\caption{
Planet torque per unit mass calculated according to eqn.~\ref{eqn:Lambda_p1} (dashed black line) and eqn.~\ref{eqn:Lambda_p2} (solid black line), and wind-driven torques per unit mass for different values of $\Sigma_{\rm A}$ such that one hemisphere of the disc has a mass flux of $5 \times 10^{-9}$ M$_{\odot}$ yr$^{-1}$ travelling through its active layer.
}
\label{fig:TorqueProfiles}
\end{figure}

We now consider the competition between wind-driven and planet torques in our layered disc models for a planet on a fixed circular orbit. We presented a similar discussion in \cite{2022A&A...658A..32L} that differs in detail with the discussion presented below, but which reached very similar conclusions. For the layered models the torque balance only needs to be considered in the active layer, and the equivalent expression to eqn.~(\ref{eqn:mdot-2pi+planet}) can be written as
\begin{equation}
-\frac{{\dot m}}{2\pi} =   \left(\frac{\partial}{\partial R} \left(R^2 \Omega\right) \right)^{-1} \left(\Lambda_{\rm p} + \Gamma_{\rm wind} \right) R \Sigma_{\rm A},
\label{eqn:mdot-Wind+Planet}
\end{equation}
where $\Gamma_{\rm wind}$ is the wind-driven torque per unit mass corresponding to eqn.~(\ref{eqn:f_wind}):
\begin{equation}
\Gamma_{\rm wind} = - \sqrt{\frac{G M_*}{R^3}} \frac{{\dot m}}{4 \pi \Sigma_{\rm A}}.
\label{eqn:Gamma_wind}
\end{equation}
We expect the planet to form a gap in the dead zone, and the radial mass flux passing through the gap and approaching the planet will depend on how $\Lambda_{\rm p}$ compares with $\Gamma_{\rm wind}$, noting that in our model the wind-driven torque applied within the active column is constant and does not change with time for a given value of $\Sigma_{\rm A}$. For $|\Lambda_{\rm p}| > |\Gamma_{\rm wind}|$ the mass flux should be blocked by the planet, and the accretion rate onto the planet should be much smaller than the mass flux through the unperturbed disc. Unlike in the viscous case, here we do not expect an increase in the mass flow into the gap as mass builds up near the gap edge. This is because we assume that $\Sigma_A$ is constant, and the torque acting in the active layer is constant, so any gas that accumulates near the gap edge will just join the dead zone, and hence will not flow towards the planet.

In the limit ($|\Lambda_{\rm p}| \ll |\Gamma_{\rm wind}|$), mass should flow relatively unimpeded towards the planet, which should be able to accrete at close to the mass flow rate through the unperturbed disc. The transition between these behaviours should occur when $|\Lambda_{\rm p}| \lesssim |\Gamma_{\rm wind}|$, and the accretion rate should be smaller than the mass flow rate through the unperturbed disc, but the accretion flow should not be completely blocked.

\begin{figure*}[htp]
\includegraphics[width=\textwidth]{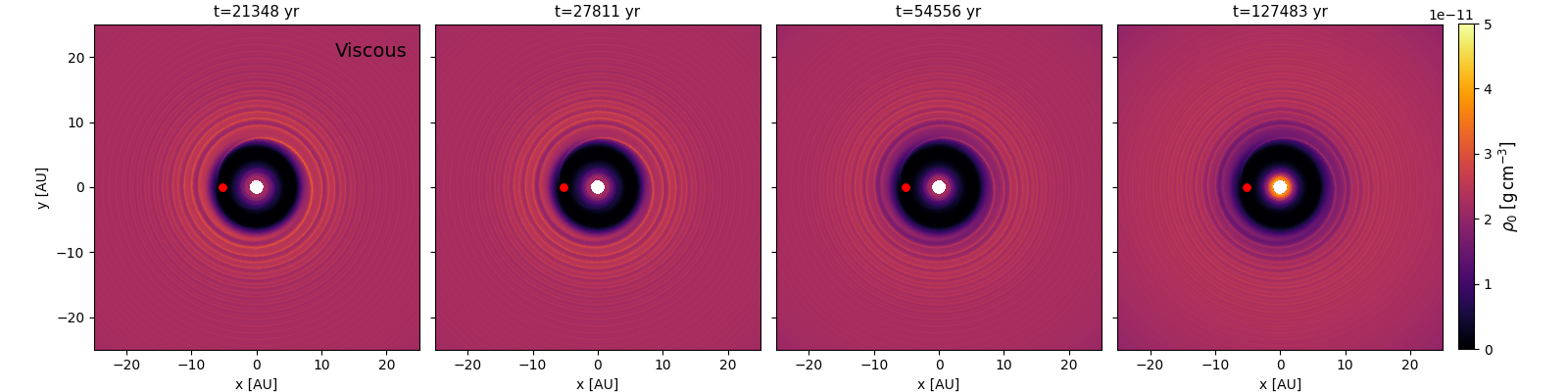}
\includegraphics[width=\textwidth]{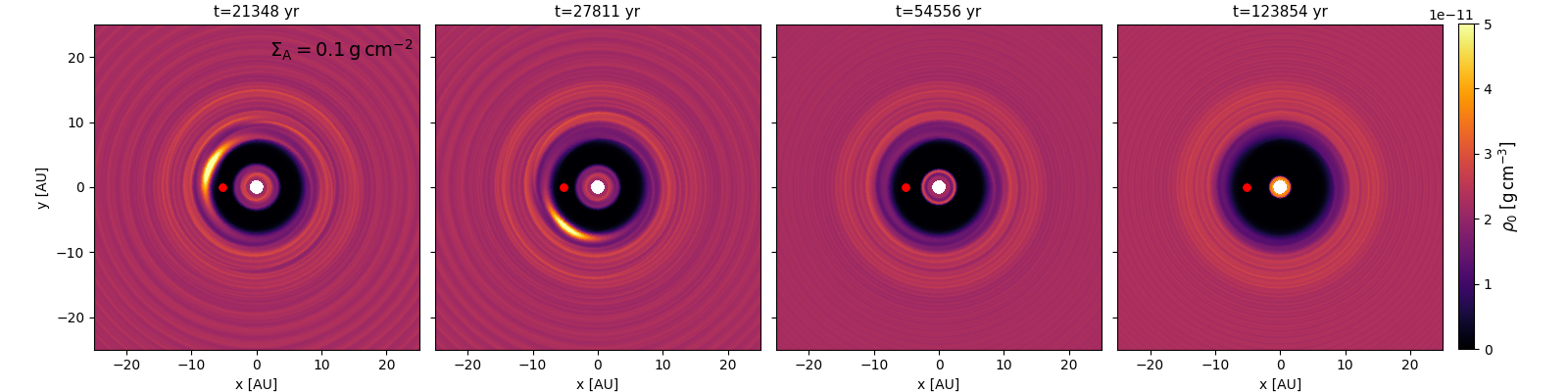}
\includegraphics[width=\textwidth]{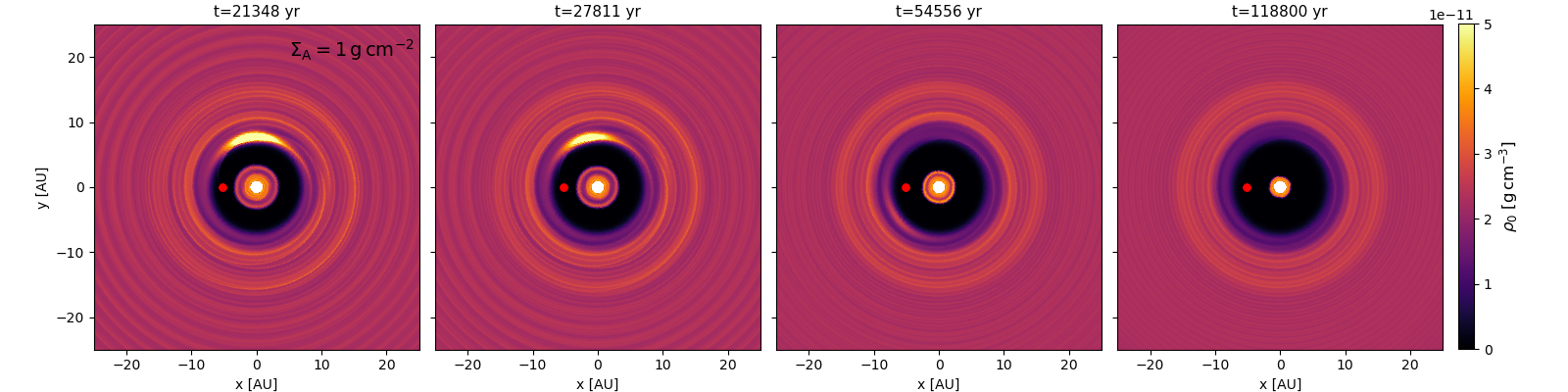}
\includegraphics[width=\textwidth]{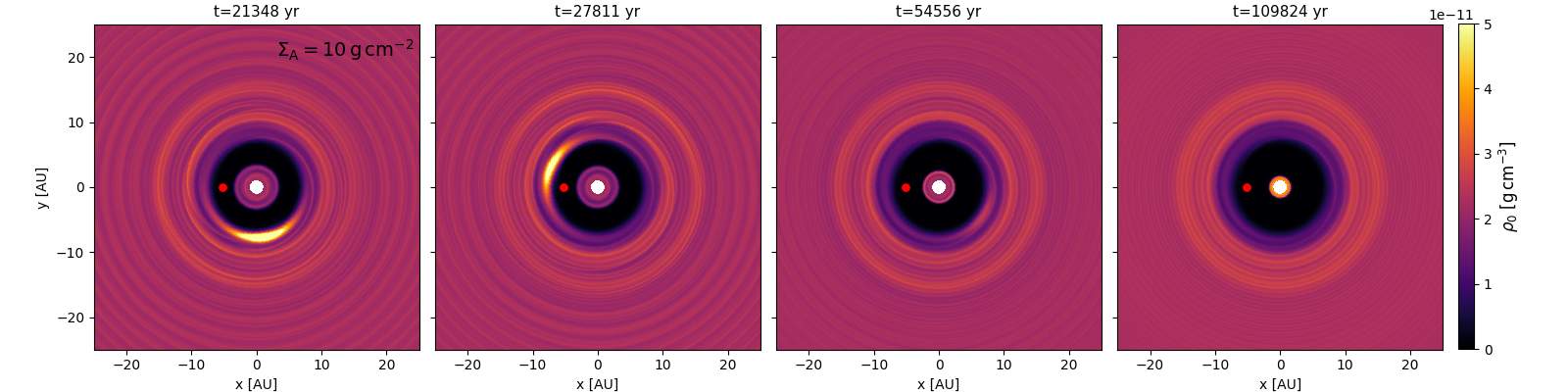}
\includegraphics[width=\textwidth]{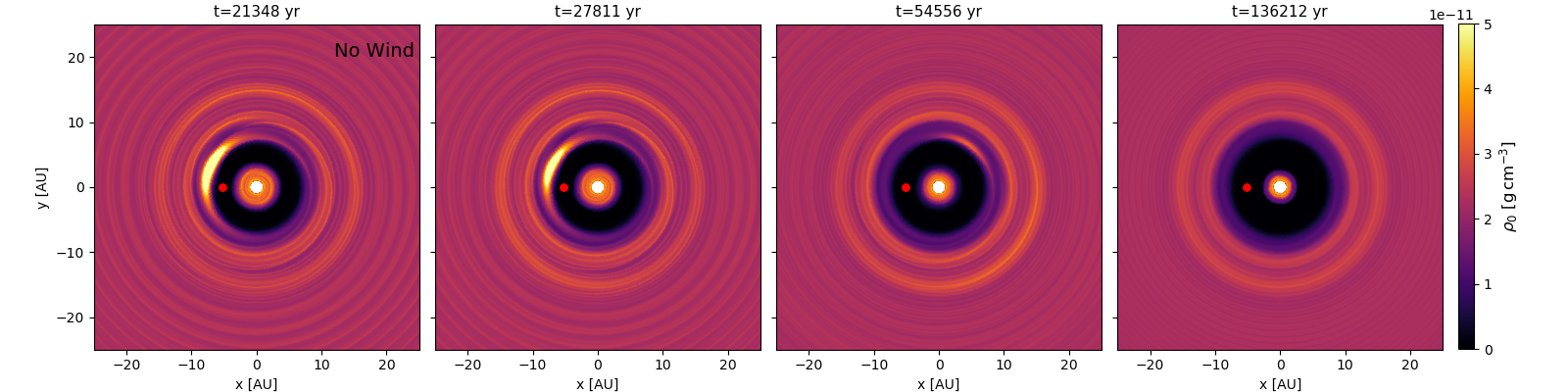}
\caption{
Contours showing the evolution of the midplane density for each of the models.}
\label{fig:density}
\end{figure*}

An expression for $\Lambda_{\rm p}$ obtained from the impulse approximation \citep{1986ApJ...309..846L} is often used in studies of disc-planet interactions, and takes the following form when the planet Hill radius exceeds the pressure scale height
\begin{equation}
\Lambda_{\rm p}=
\begin{cases}
{\rm sign}(R-R_{\rm p}) \frac{8}{9 \pi} \frac{q_{\rm p}^2 G M}{2 R_{\rm p}} \left( \frac{R_{\rm p}}{\Delta_{\rm p}} \right)^4, & \text{if } |R - R_{\rm p}| > R_{\rm H} \\ \\
\,\,\,\,\,\,\,\,\,\,\,\,\,\, \frac{(R-R_{\rm p})}{R_{\rm H}} \frac{8}{9 \pi} \frac{q_{\rm p}^2 G M}{2 R_{\rm p}} \left( \frac{R_{\rm p}}{\Delta_{\rm p}} \right)^4, & \text{otherwise}
\end{cases}
\label{eqn:Lambda_p1}
\end{equation}
where $\Delta_{\rm p} = {\rm Max}\left( R_{\rm H}, R-R_{\rm p}\right)$, $q_{\rm p}=M_{\rm p}/M_*$, and $R_{\rm H}$ is the Hill radius. The second term in eqn.~(\ref{eqn:Lambda_p1}) is a modification that ensures $\Lambda_{\rm p}$ is continuous as it passes through the planet's location. Using 3D hydrodynamical simulations of disc-planet interactions similar to those presented here, \cite{2010ApJ...724..730D} have examined how $\Lambda_{\rm p}$ varies in different disc models and for different planet masses, and they provide the following expression based on fits to their simulation results (so the expression agrees much more closely with hydrodynamical simulations than does eqn.~\ref{eqn:Lambda_p1}):
\begin{equation}
\Lambda_{\rm p} = \frac{1}{2}{\cal F}(x,p,q) \Omega_{\rm p}^2 R_{\rm p}^2 q_{\rm p}^2 \left(\frac{R_{\rm p}}{R_{\rm H}}\right)^4,
\label{eqn:Lambda_p2}
\end{equation}
where 
\begin{equation}
{\cal F} (x,p,q)=\left( p_1 e^{\left( - \frac{(x+p_2)^2}{p_3^2} \right)} + p_4 e^{\left( - \frac{(x - p_5)^2}{p_6^2} \right)} \right) \times \tanh{(p_7 - p_8 x)}.
\label{eqn:F}
\end{equation}
We note that the values $p$ and $q$ in ${\cal F}(x,p,q)$ refer to the power-law indices that define the density and temperature profiles in eqns.~(\ref{eqn:rho}) and (\ref{eqn:T}), and the factor of $1/2$ in eqn.~(\ref{eqn:Lambda_p2})  should only be included to account for non-linear effects when gap opening planets are considered. Similarly in this limit we have $x=(R-R_{\rm p})/R_{\rm H}$. The constants $p_1$, $p_2$, etc. in eqn.~(\ref{eqn:F}) depend on the disc structure, and for the power-law indices $p=-1.5$ and $q=-1$ we consider in this work the values of these constants are given in table~1 of \cite{2010ApJ...724..730D}. 

\begin{figure}
\includegraphics[width=\columnwidth]{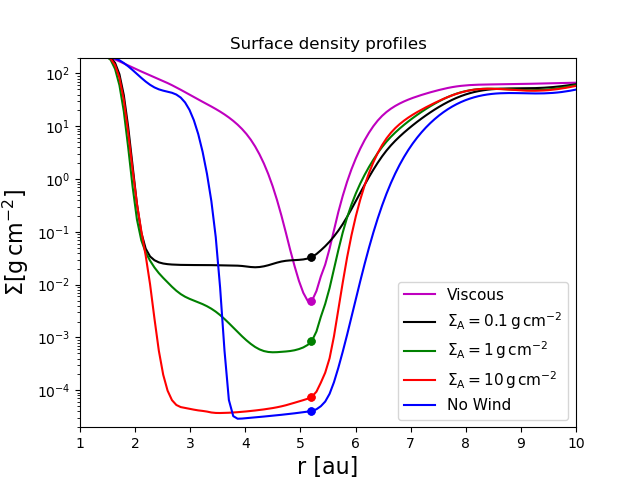}
\caption{
Azimuthally averaged surface density profiles from all disc models computed once each run had evolved for $\sim$49,000 yr after the value of $K$ was increased from 6 to 30 in eqn.~\ref{eqn:dmpdt}. The planet positions are indicated by the filled circles. Note that the inner boundary condition influences the surface density profiles close to the inner boundary. The profiles shown for each run correspond to the right-hand-most midplane density contours shown in fig.~\ref{fig:density}.
}
\label{fig:SigmaProfiles}
\end{figure}

Figure~\ref{fig:TorqueProfiles} shows the values of $\Lambda_{\rm p}$ from eqns.~(\ref{eqn:Lambda_p1}) and (\ref{eqn:Lambda_p2}), and also the values of $\Gamma_{\rm wind}$ for each of the layered disc models. It is clear that eqn.~(\ref{eqn:Lambda_p1}) predicts torque values that are too large. Comparing the values from  eqn.~(\ref{eqn:Lambda_p2}) with the wind-driven torques for the different disc models, we expect the accretion rate onto the planet to be close to the mass flow rate through the unperturbed disc for $\Sigma_{\rm A}=0.1$~g~cm$^{-2}$. For $\Sigma_{\rm A}=1$~g~cm$^{-2}$, we expect the accretion rate to be moderately impeded, and for $\Sigma_{\rm A}=10$~g~cm$^{-2}$ we expect the accretion flow towards the planet to be blocked, and hence for there to be a very significant reduction in the accretion rate onto the planet compared to the unperturbed mass flux through the disc.

\section{Results}
\label{sec:Results}
\subsection{Disc structure}
Figure~\ref{fig:density} shows the evolution of the midplane density in each of the models, and a common feature is the formation of a deep annular gap around the planet's orbital location. As described in our earlier work \citep{2021A&A...646A.166L, 2022A&A...658A..32L}, the low viscosity models all show the development of a strong vortex that grows via the Rossby Wave Instability at the density maximum associated with the edge of the gap \citep{2006MNRAS.370..529D, 1999ApJ...513..805L}. The vortex is a source of strong spiral waves that propagate outwards and dissipate, forming a secondary gap and a secondary density maximum that can be observed as a high density ring just exterior to 10~au in fig.~\ref{fig:density}. The vortices migrate inwards and dissipate as they move away from the pressure bump, as can be seen in the rightmost panels of the figure, and the low viscosity employed in these models means that the ring and gap features generated by the vortex survive over long timescales. These models maintain long memories of earlier evolution. As expected, the Viscous model does not show the development of a long lived vortex or the secondary features that might arise from one, and instead the disc exterior to the planet has a smooth structure on which spiral density waves are superposed.

The azimuthally averaged surface density profiles for each of the models are shown in fig.~\ref{fig:SigmaProfiles}. Here we see significant variation between the models. The gap for the Viscous model remains relatively narrow, since the formation of the gap causes the viscous flow of material into the gap from both the inner and outer disc. The No Wind model shows the formation of a much deeper and wider gap than the Viscous model due to the very low viscosity adopted in this run, allowing the planet to more strongly repel material on both sides of its orbit. The inner-outer disc asymmetry observed in this run is likely due to the presence of a vortex at the outer edge of the gap during most of the simulation. The spiral waves emitted by this, and its eventual dissipation, would have provided a source of mass flow towards the planet at the outer gap edge.

The models with wind-driven accretion flows all show the same strongly asymmetrical gap structure, where the asymmetry this time is significantly larger than in the No Wind case. This arises because the wind-driven torque drives the disc gas towards the planet in the outer disc, opposite to the direction that the planet torque tries to drive material, and away from the planet and towards the star in the inner disc, where the planet torque acts in the same direction as the wind-driven torque. We note that keeping the planet on a fixed orbit rather than allowing it to migrate somewhat exaggerates the asymmetry of the gap, as shown by the results in \cite{2022A&A...658A..32L}. Allowing the planet to accrete gas efficiently also contributes to this asymmetry, by blocking the flow from the outer to the inner disc even in models where the wind-driven flow is not strongly impeded by the tidal torque from the planet. The $\Sigma_{\rm A}=0.1$~g~cm$^{-2}$ run has the shallowest gap, showing that that gas is able to flow into the gap region and is not strongly impeded by the planet. The $\Sigma_{\rm A}=10$ case, however, displays a very deep gap indicating that gas flow into the gap is strongly diminished by the planet. The $\Sigma_{\rm A}=1$ case is intermediate between these two, showing it lies in the transition between weakly and strongly impeded radial gas flow. Finally, it is worth noting that an interesting implication of these different disc structures, all arising from the same global accretion rate through the disc, is that accurately inferring planet masses from observations of gap structures in protoplanetary discs becomes much more difficult if accretion through the disc is driven by an unknown combination of stresses originating from turbulent viscosity and a magnetised wind.

\subsection{Gas accretion rates}
\begin{figure*}
\includegraphics[width=\columnwidth]{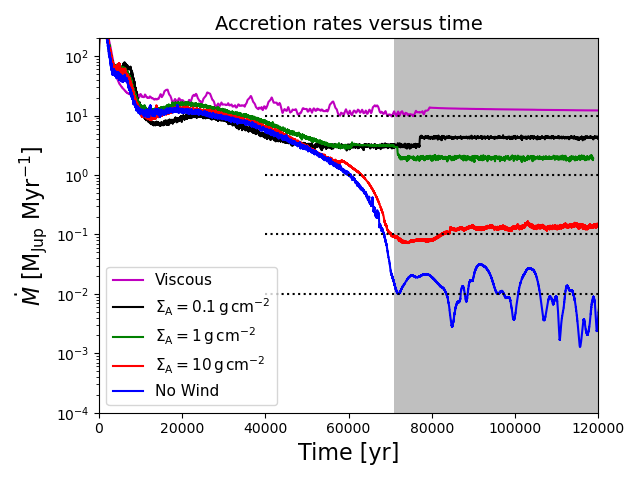}
\includegraphics[width=\columnwidth]{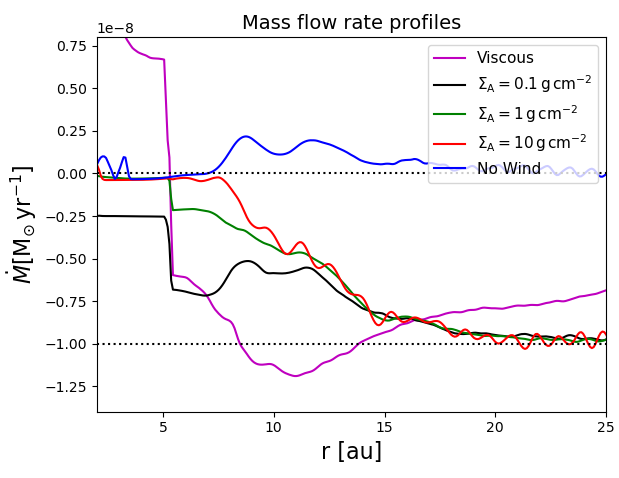}
\caption{
Gas accretion and radial mass flow rates. Left panel: Gas accretion rates onto the planets versus time for all disc models. Moving from top to bottom, the horizontal black dotted lines correspond to accretion rates of 10, 1, 0.1, and 0.01 M$_{\rm Jup}$ Myr$^{-1}$. The grey shaded region highlights the temporal domain within which the value of $K$ in eqn.~\ref{eqn:dmpdt} is increased from 6 to 30. For the Viscous run and those with different values of $\Sigma_{\rm A}$, the increase coincides with the discontinuous changes to the accretion rates that can be observed. The value of $K$ for the No Wind case was increased at time $\sim 87,000$~yr. Right panel: Radial mass flux profiles for all disc models containing Jovian mass planets. The upper and lower horizontal black dotted lines correspond to radial mass fluxes of 0 and 10$^{-8}$~M$_{\odot}$ yr$^{-1}$, respectively. The mass fluxes were computed at the times corresponding to the surface density profiles displayed in fig.~\ref{fig:SigmaProfiles}.
}
\label{fig:mdot-Mp300}
\end{figure*}
The accretion rates onto the planets in each model are shown in the left panel of figure~\ref{fig:mdot-Mp300}, and the results are in line with the expectations described in Section~\ref{sec:Theory}. The Viscous model shows a steady accretion rate corresponding to ${\dot M_{\rm p}} \simeq 12$~M$_{\rm Jup}$~Myr$^{-1}$, such that a Jovian mass planet will double its mass in $\lesssim 10^5$~yr. Time-averaged radial profiles of the vertically and azimuthally integrated mass fluxes through the discs with embedded planets are shown in the right panel of figure~\ref{fig:mdot-Mp300}. Exterior to the planet, the viscous disc has a mass flux towards the planet close to the unperturbed value. The disc interior to the planet shows an outward mass flow, again towards the planet. Hence, the planet at $r_{\rm p}=5.2$~au is being fed from both sides, and the jump in the mass flux rate at the planet location corresponds to the steady accretion rate recorded in the left panel.

The planet in the $\Sigma_{\rm A}=0.1$ model has a steady accretion rate of ${\dot M_{\rm p}} \simeq 6$~M$_{\rm Jup}$ Myr$^{-1}$, giving a mass doubling time of $2\times 10^5$ yr. The mass flux far from the planet in the right panel is again close to the unperturbed value, and decreases somewhat closer to the planet. The jump in the radial mass flux at the planet location again corresponds to the accretion rate recorded in the left panel, showing that the gas supplied to the planet by the wind-driven torque is largely accreted by it, although a fraction of this gas flows past the planet and into the inner disc. 

The planet in the $\Sigma_{\rm A}=1$ model sustains an accretion rate of ${\dot M}_{\rm p} \simeq 2$~M$_{\rm Jup}$ Myr$^{-1}$, giving a mass doubling time of $\simeq 5 \times 10^5$~yr. The right panel of figure~\ref{fig:mdot-Mp300} shows a similar picture to the $\Sigma_{\rm A}=0.1$ model, with the mass flux at large radius being equal to the unperturbed value and decreasing somewhat at radii closer to the planet. The jump in the mass flux at the planet location again corresponds to the accretion rate displayed in the left panel. 

The $\Sigma_{\rm A}=10$ runs shows the greatest influence of the planet torque on the gas accretion rate. Here ${\dot M}_{\rm p} \simeq 10^{-1}$~M$_{\rm Jup}$ Myr$^{-1}$, giving a mass doubling time of 10~Myr. In agreement with the discussion in Section~\ref{sec:Theory}, the wind-driven radial gas flow towards the planet is almost completely blocked by the planet torque, giving rise to the dramatically reduced accretion rate. 

Finally, the No Wind model shows an averaged accretion rate of ${\dot M_{\rm p}} \sim 10^{-2}$M$_{\rm Jup}$ Myr$^{-1}$. This small accretion rate indicates that the inclusion of a small viscosity in the wind-driven models does not significantly contribute to the gas flow towards the planets. The right panel of figure~\ref{fig:mdot-Mp300} shows the spiral waves launched by the planet into the outer disc cause a modest outward flow of mass in the No Wind model, and beyond 20~au the disc shows no net flow of mass as expected. Closer to the planet in the gap region, there is a very weak inflow towards the planet which arises because of the small imposed viscosity acting within the very deep gap.

\begin{figure*}
\includegraphics[width=0.33\textwidth]{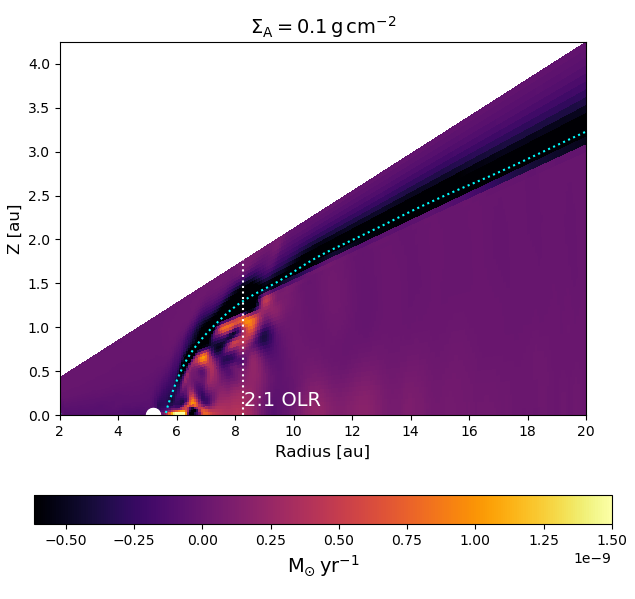}
\includegraphics[width=0.33\textwidth]{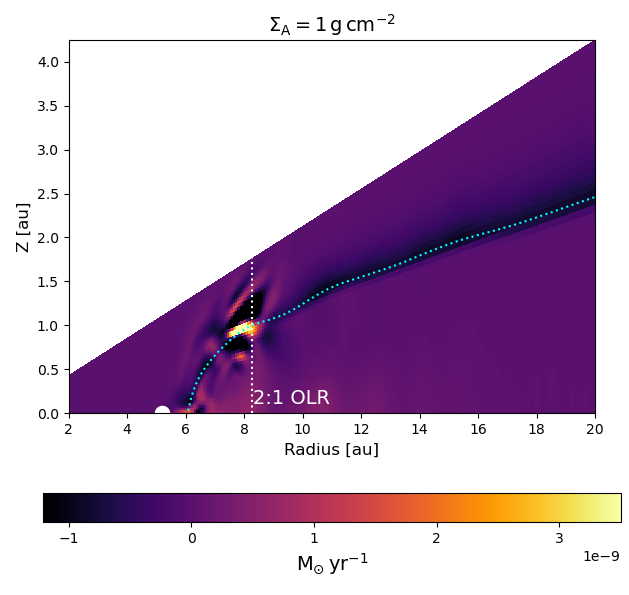}
\includegraphics[width=0.33\textwidth]{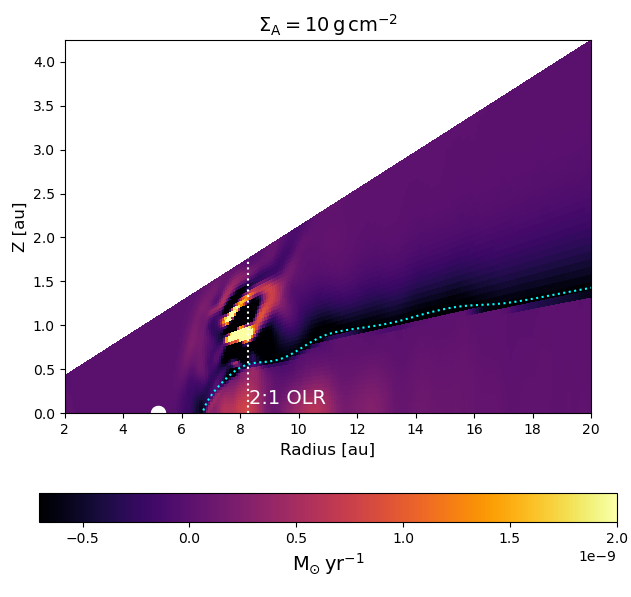}
\caption{
Contour plots showing time-averaged and azimuthally-integrated radial mass flux distributions in the various layered disc models with embedded planets. The cyan dotted line indicates the location where the column density measured from the disc surface corresponds to $\Sigma_{\rm A}$. The vertical white dotted line denotes the location of the 2:1 outer Lindblad resonance. Note that only the upper hemisphere of the disc is simulated.}
\label{fig:MfluxContoursWithPlanet}
\end{figure*}

Figure~\ref{fig:MfluxContoursWithPlanet} shows time averages of the azimuthally integrated radial mass fluxes in the three simulations with wind-driven accretion flows. The time averages were computed using 100 snapshots that were output with an interval of one planet orbit, equivalent to 11.86~yr, starting at times slightly before the times that correspond to the surface density profiles displayed in fig.~\ref{fig:SigmaProfiles}. The panels show the accretion flow being largely confined to a narrow region in the vertical direction close to the transition between the active and dead zones, a feature which persists all the way into the gap. It is noteworthy that in all models the full column density, $\Sigma(0)$, falls below $\Sigma_{\rm A}$ in the gap, such that the integrated accretion flow must fall below its unperturbed value there because the wind-driven torque per unit mass applied to the active column is constant in the wind model we have adopted. Figure~\ref{fig:MfluxContoursWithPlanet} illustrates the fact that in 3D hydrodynamical simulations the flow is considerably more complex than suggested by the simple 1D picture presented in Section~\ref{sec:Theory}, and the influence of resonances such as the 2:1 outer Lindblad resonance induce spatial and temporal variability in the flow. Nonetheless, the simple picture of the long term net accretion flow onto the planet being determined by the balance between the applied wind-driven torque and the planet tidal torque seems to hold, and explains the behaviour observed in the simulations as a function of $\Sigma_{\rm A}$ and the applied viscosity.

\section{Discussion}
\label{sec:Discussion}
\subsection{Combined migration and gas accretion rates}
In \cite{2022A&A...658A..32L} we presented simulations of migrating and non-accreting Jovian mass planets embedded in layered disc models in which the total mass flux towards the star is ${\dot m}=10^{-8}$~M$_{\odot}$~yr$^{-1}$. Here we have presented simulations of accreting and non-migrating Jovian mass planets in similar layered models. In each of these scenarios, the behaviour is determined by the ability or otherwise of the planetary torque to block the wind-induced mass flow occurring in the active layer. In both studies we observe a dramatic change in behaviour when considering the two extreme cases simulated, $\Sigma_{\rm A} =0.1$~g~cm$^{-2}$ and $\Sigma_{\rm A}=10$~g~cm$^{-2}$, with the transition in behaviour occurring between these values.

For migrating planets, \cite{2022A&A...658A..32L} show that once the vortex that forms in the layered disc models has dissipated, the long term migration behaviour depends on the mass flux through the gap. When the gas flows unimpeded through the gap then we can consider the disc to consist of two distinct parts: an inert dead zone that contains most of the mass, and a low mass active zone that sustains a constant mass flux at all radii. The planet forms a gap in the dead zone and sits in a location that minimises the net torque arising from its interaction with the inner and outer disc. The flow of gas through the gap maintains only a small amount of mass in the gap that barely influences the migration rate. Hence in this case we have very slow inward migration of the planet. The results in this paper show that this slow mode of migration will be accompanied by rapid gas accretion, since an unimpeded accretion flow reaches the planet. We note, however, that this conclusion depends on the efficiency with which the planet can accrete the gas that is supplied to it. As we discuss below, there is considerably uncertainty about what this efficiency is.

If we consider what would happen when a planet is undergoing slow migration and is able to accrete gas at close to the supply rate, then the flow from the outer to the inner disc would be interrupted. Over time the wind-driven torque acting on the inner disc would deplete it, and this would have the effect of modifying the balance of Lindblad torques acting on the planet through interaction with the inner and outer disc. This is one way in which allowing a planet to accrete gas while it migrates could in principle modify the migration behaviour compared to a scenario in which the planet does not accrete gas. We speculate that in this scenario the planet would migrate away from the outer gap edge until the torque it experiences from it becomes very small, at which point its migration would stall. Gas accretion, however, would continue to occur, and hence there seems to be little prospect of halting accretion and migration simultaneously. Simulations that explore this scenario will be presented in a forthcoming publication.

When the gas flow is blocked by the planet torque, then the migration picture changes.  As the planet migrates inwards and away from the outer edge of the gap, the laminar accretion flow fills in the gap behind the planet. Hence, migration is sustained by the gas inflow and the migration rate is determined by the rate at which the gap is refilled, leading to the following estimate for the migration rate
\begin{equation}
    {\dot r}_{\rm {p}} = \frac{\dot M}{2 \pi r_{\rm p} \Sigma},
    \label{r_p_dot}
\end{equation}
where $\Sigma$ is the total surface density at the edge of the gap.
For a planet at $r_{\rm p}=5.2$~au, the simulations in \cite{2022A&A...658A..32L} show the migration speed is $\sim 15$~au~Myr$^{-1}$ when the accretion flow through the disc is $10^{-8}$~M$_{\odot}$~yr$^{-1}$ and $\Sigma_{\rm A}=10$~g~cm$^{-2}$. The results in this paper show that the gas accretion rate will be very small because the rate at which gas is supplied to the planet is ${\dot m} \sim 0.1$~M$_{\rm Jup}$~yr$^{-1}$ while the planet is in this faster mode of migration.

To summarise, a Jovian mass planet that can efficiently accrete any gas that is supplied to it, embedded at $r_{\rm p}=5.2$~au in a disc with a wind-induced accretion flow of $10^{-8}$~M$_{\odot}$~yr$^{-1}$, will undergo slow migration at a speed ${\dot r}_{\rm p} \sim 3$~au~Myr$^{-1}$, and rapid gas accretion at a rate of ${\dot M}_{\rm p} \sim 5$~M$_{\rm Jup}$~Myr$^{-1}$ if $\Sigma_{\rm A} = 0.1$~g~cm$^{-2}$. If $\Sigma_{\rm A}=10$~g~cm$^{-2}$, however, then the migration speed will be faster ${\dot r}_{\rm p} \sim 15$~au~Myr$^{-1}$, and the accretion rate will reduce to ${\dot M}_{\rm p} \lesssim 0.1$~M$_{\rm Jup}$~Myr$^{-1}$. Migration and gas accretion rates that are intermediate between these values will be obtained for values of $\Sigma_{\rm A}$ that lie between these limiting cases, as demonstrated here and in \cite{2022A&A...658A..32L} for the case of $\Sigma_{\rm A}=1$~g~cm$^{-2}$.

\begin{figure}
\includegraphics[width=\columnwidth]{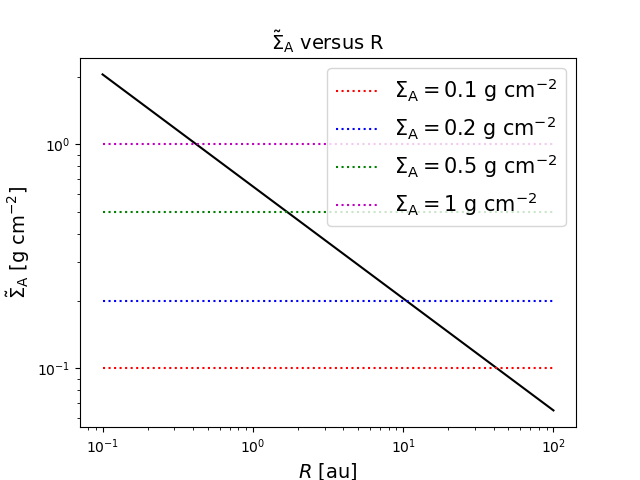}
\caption{${\tilde \Sigma_{\rm A}}$ versus radius for a Jovian mass planet located in a disc with ${\dot m}=10^{-8}$~M$_{\odot}$~yr$^{-1}$. Also shown are various values of $\Sigma_{\rm A}$, indicating where an embedded planet would transition between fast migration and slow accretion to slow migration and fast accretion.
}
\label{fig:SigmaCrit}
\end{figure}
\subsection{Behaviour as a function of orbital radius}
We have considered gas accretion onto Jovian mass planets orbiting at $r_{\rm p}=5.2$~au in layered disc models. Here we consider how the evolution depends on orbital radius, assuming that $\Sigma_{\rm A}$ varies weakly with stellocentric distance. As discussed in \cite{2022A&A...658A..32L} and Sect.~\ref{sec:Theory}, the transition between rapid and slow migration (and hence between slow and rapid gas accretion) occurs when $\Lambda_{\rm p} \sim \Gamma_{\rm wind}$. Using eqns.~\ref{eqn:Lambda_p2} and \ref{eqn:Gamma_wind} we can determine the critical value of $\Sigma_{\rm A}$ that corresponds to this transition in behaviour. 
We note that the function ${\cal F}$ defined in eqn.~(\ref{eqn:F}) has a maximum value ${\cal F}_{\rm max} \sim 0.02$. The critical value, ${\tilde \Sigma_{\rm A}}$, can be written
\begin{equation}
{\tilde \Sigma_{\rm A}} = 0.285 \, {\rm g} \, {\rm cm}^{-2} \left(\frac{\dot m}{10^{-8} \, {\rm M}_{\odot} \, {\rm yr}^{-1}} \right) \left(\frac{M_{\rm p}}{M_{\rm Jup}} \right)^{-2/3} \left(\frac{r_{\rm p}}{5.2 \, {\rm au}} \right)^{-1/2},
\label{Sigma_crit}
\end{equation} 
where ${\dot m}$ corresponds to the total mass flux through the disc due to the wind-driven torque. When $\Sigma_{\rm A} > \tilde \Sigma_{\rm A}$ the accretion flow will be blocked and we expect fast migration, and when $\Sigma_{\rm A} < \tilde \Sigma_{\rm A}$ the accretion flow will not be blocked and migration will be slow. 

As remarked upon in \cite{2022A&A...658A..32L}, the $r_{\rm p}^{-1/2}$ dependence of ${\tilde \Sigma_{\rm A}}$ means that a migrating planet can transition between the fast and slow modes of migration as it migrates inwards. Figure~\ref{fig:SigmaCrit} shows how ${\tilde \Sigma_{\rm A}}$ varies with radius in a disc with ${\dot m}=10^{-8}$~M$_{\odot}$~yr$^{-1}$ containing a Jovian mass planet, and shows the transition radius (where the lines cross in the figure) between fast and slow migration for different values of $\Sigma_{\rm A}$. An important issue is determining an appropriate value for $\Sigma_{\rm A}$. Although X-rays and cosmic rays are important sources of ionisation in protoplanetary discs, and penetrate into columns of depth $\sim 10$~g~cm$^{-2}$ and 100~g~cm$^{-2}$, respectively, the strong magnetic coupling in the surface layers of protoplanetary discs that allows a magnetised wind to be launched is thought to depend on the ionisation of sulphur and carbon atoms by UV photons from the star \citep{2011ApJ...735....8P}. The column density of this layer is such that we expect $\Sigma_{\rm A} < 1$~g~cm$^{-2}$. As shown by figure~\ref{fig:SigmaCrit}, values of $0.1 \le \Sigma_{\rm A} \le 1$~g~cm$^{-2}$ give rise to locations where the planet transitions between fast and slow migration (and vice versa for gas accretion) that cover a very wide range of radii in protoplanetary discs, from $r\sim 50$~au when $\Sigma_{\rm A}=0.1$ to $r\sim 0.4$~au when $\Sigma_{\rm A}=1$.

If we suppose the disc has $\Sigma_{\rm A} =0.5$~g~cm$^{-2}$ at all disc radii, and a Jovian mass planet starts migrating at $r_{\rm p}=5.2$~au, then $\Sigma_{\rm A} > {\tilde \Sigma_{\rm A}}(r=5.2~{\rm au})$ and migration should be in the fast regime with gas accretion being in the slow regime. As the planet migrates inwards ${\tilde \Sigma_{\rm A}}$ increases, and at $r_{\rm p} \sim 1.6$~au we have $\Sigma_{\rm A} = {\tilde \Sigma_{\rm A}}$, so that migration should start to enter the slow regime and gas accretion should transition to the rapid regime. This evolutionary scenario nicely explains the large number of giant planets observed with orbital periods $P\gtrsim 100$~days, but also suggests that giant planet masses should on average show an inverse relationship with orbital period near the transition radius.

Figure~\ref{fig:RVdata} displays mass-period diagrams for giant planets discovered by radial velocity surveys {(top panel is for all planets, bottom panel is for low eccentricity planets with $e < 0.3$ that are assumed to have not undergone strong gravitational scattering after removal of the gas disc )\footnote{Data downloaded from https://exoplanetarchive.ipac.caltech.edu on 26 May 2022}, and they do not show strong evidence for a significant increase in planet masses at orbital periods less than a few hundred days, independent of the eccentricity range being considered. A local minimum can be observed at $\sim 400$~days, but this does not seem to be statistically significant, and is not consistent with the idea that Jovian mass planets that form at $\sim 5.2$~au, migrate inwards rapidly, while slowly increasing their masses, and then slow down their migration and significantly increase their gas accretion rates as they reach orbital radii $\sim 1$~au.

\begin{figure}
\includegraphics[width=\columnwidth]{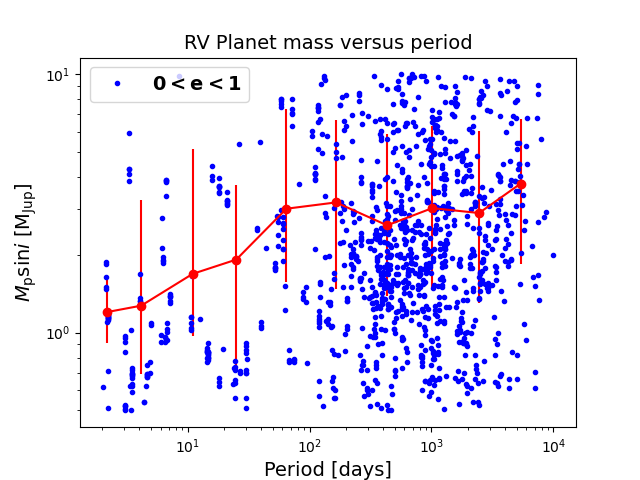} \\
\includegraphics[width=\columnwidth]{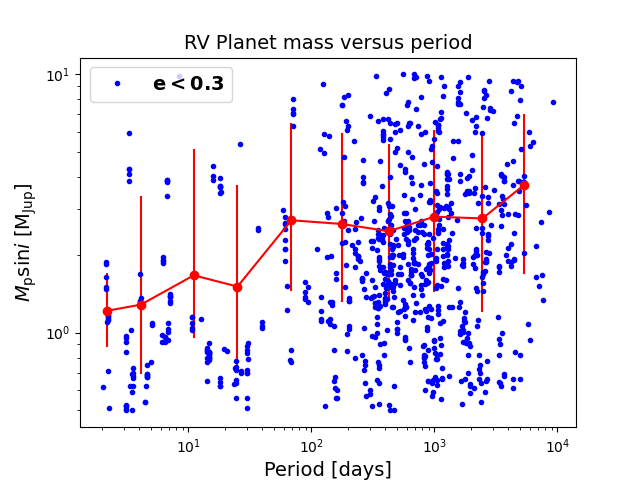}
\caption{$M_{\rm p} \sin{i}$ versus period for exoplanets discovered by radial velocity surveys. Top panel is for all planets, bottom panel is for low eccentricity planets where $e < 0.3$. Red points show the average masses of planets binned in radius, where the bin width is 0.4 in $\log_{10}{P}$. Error bars are for one standard deviation of the planet mass distribution in each bin.}
\label{fig:RVdata}
\end{figure}

\subsection{Gas supply rate versus accretion rate}
\label{sec:GasSupply}
The discussion in the previous subsection suggests that giant planets somehow avoid accreting gas at the rate supplied by the disc over significant timescales, but instead either accrete for only relatively short time periods because they form towards the end of the disc lifetime, or accrete the supplied gas at a significantly lower rate than assumed in simulations such as the ones presented here. 

There have been a number of recent high resolution 3D studies of the gas flows within the Hill spheres and surrounding regions for planets of various masses embedded at different radii in protoplanetary discs \citep[e.g.][]{2019ApJ...887..152F, 2019A&A...630A..82L, 2016MNRAS.460.2853S, 2021A&A...646L..11M, 2022arXiv220211422M}. These studies have often utilised radiation-hydrodynamics codes, and hence are able to examine the cooling of the gas and the rate at which it accretes onto the planet. \cite{2021A&A...646L..11M} and \cite{2022arXiv220211422M} examined gas accretion onto cores ranging in mass between 1 and 10~M$_{\oplus}$, orbiting close to the star at $0.1$~au, and showed that the cooling and contraction of the planet envelope can be switched off completely by the continuous advection of high entropy gas into the Hill sphere from the surrounding disc. The advection occurs at a rate that scales with the local orbital period, and so while this is likely to be an important effect for planets close to the star, it is not clear that it will remain so at large stellocentric distances. This point is supported by the 3D radiation-hydrodynamic simulations presented by \cite{2019A&A...630A..82L}, who showed that the advection of gas into the Hill sphere is still an important phenomenon for planets orbiting at 5.2~au, but that Jovian mass planets orbiting there accrete gas at a rate that leads to the mass doubling time being $\lesssim 10^{5}$~yr. Hence, the conclusion of that study was that thermal effects occurring within the Hill sphere are unlikely to slow down the accretion of gas if it is supplied by the disc at a rate of ${\dot m} \sim 10^{-8}$~M$_{\odot}$~yr$^{-1}$. Other effects might be important in slowing down the accretion of gas once it has been supplied to the vicinity of the planet, for example magnetic effects occurring in the Hill sphere, such as considered by \cite{2013ApJ...779...59G}, or an increase in the magnetic torque in the gap leading to a faster radial flow of gas past the planet, such that it accretes at slower rate. At the present time, however, we do not have evidence of the efficacy of these effects in slowing down accretion, and it is difficult to escape the conclusion that the final masses of most giant planets are determined by the lengths of time that they are present within their protoplanetary discs, rather than by a physical process that acts to slow the accretion of gas that is supplied by the protoplanetary disc, or a process that slows the supply to the planet's location deep inside the Hill sphere.

\section{Conclusions}
\label{sec:Conclusions}
We have presented the results of 3D hydrodynamical simulations of accreting giant planets embedded in protoplanetary discs that globally sustain accretion flows towards the star of ${\dot m} = 10^{-8}$~M$_{\odot}$~yr$^{-1}$. We consider a classical $\alpha$ viscous disc model, and a sequence of very low viscosity models in which a laminar accretion flow is driven in the surface layers of the disc by an external torque, that is supposed to mimic the effects of a magnetocentrifugally driven wind. In these latter models, the column density of the actively accreting layer is varied in the range $\Sigma_{\rm A}=0.1$ to 10~g~cm$^{-2}$, such that the radial speed, and the torque per unit mass acting on the accreting gas, increases as the column density of the actively accreting gas decreases, in order to maintain the given accretion rate.

The main result obtained is that the accretion rate onto the planet varies significantly between the models, in spite of the fact that the accretion rate supplied through the disc is always ${\dot m} = 10^{-8}$~M$_{\odot}$~yr$^{-1}$. The viscous model produces an accretion rate that leads to a mass doubling time for a Jovian mass planet of $\sim 10^5$~yr (0.1 Myr), in agreement with numerous previous studies. The wind-driven models, however, produce accretion rates such that the mass doubling times vary between 0.2 and 10~Myr for $\Sigma_{\rm A}=0.1$ and 10~g~cm$^{-2}$, respectively. 

Fundamentally, this result arises because viscous discs adjust their accretion flows in the presence of a planet-induced gap, such that a large accretion rate through the gap and onto the planet can be maintained. The model for the wind-induced torque we have adopted, on the other hand, assumes the torque per unit mass acting on the accreting gas does not change in the gap regions, and hence for larger values of $\Sigma_{\rm A}$ the tidal torque from the planet can overwhelm the torque acting in the accreting layer, and effectively block the accretion flow. For small values of $\Sigma_{\rm A}$, the planet torque is smaller in magnitude than that acting in the disc surface layers, and hence the accretion flow is relatively unimpeded by the planet, such that a large accretion rate onto it can be maintained. For the parameters adopted in our models, the transition in behaviour between fast and slow accretion occurs between $0.1 < \Sigma_{\rm A} < 10$~g~cm$^{-2}$. 

In a recent study \citep{2022A&A...658A..32L}, we considered the migration of non-accreting Jovian mass planets in layered disc models, similar to those presented here. In that case, there is also a transition in behaviour such that migration is fast when $\Sigma_{\rm A}=10$~g~cm$^{-2}$ and slow when $\Sigma_{\rm A}=0.1$~g~cm$^{-2}$. Hence, we expect that planets will accrete slowly when local disc conditions allow for rapid inward migration, and will accrete rapidly when conditions cause migration to be slow (assuming no process operates that can impede accretion onto the planet of gas that is supplied by the accretion flow in the background disc). We will present the results of an ongoing study of migrating and accreting giant planets embedded in layered disc models in a forthcoming publication.

The model of the wind-driven accretion torque we have presented is highly simplified, and serves the purpose of providing a framework for understanding how giant planets evolve when embedded in non-viscous and accreting protoplanetary discs. In future work, we will improve on the simplifications of the present model by employing non-ideal MHD simulations, and examine how giant planets migrate and accrete in discs that include a more complete treatment of the relevant physics.

\begin{acknowledgements}
We thank the anonymous referee for a constructive report that helped improve the paper.
RPN acknowledges support from STFC through grants ST/P000592/1 and ST/T000341/1.
This research utilised Queen Mary's Apocrita HPC facility, supported by QMUL Research-IT (http://doi.org/10.5281/zenodo.438045). This work was performed using the DiRAC Data Intensive service at Leicester, operated by the University of Leicester IT Services, which forms part of the STFC DiRAC HPC Facility (www.dirac.ac.uk). The equipment was funded by BEIS capital funding via STFC capital grants ST/K000373/1 and ST/R002363/1 and STFC DiRAC Operations grant ST/R001014/1. DiRAC is part of the National e-Infrastructure.
EL and AM acknowledge support by DFG-ANR supported GEPARD project
 (ANR-18-CE92-0044 DFG: KL 650/31-1). We also acknowledge HPC resources from GENCI DARI n.A0120407233 and from "Mesocentre SIGAMM" hosted by Observatoire de la C\^ote d'Azur. AM acknowledges support from the ERC project HolyEarth - 101019380. EL wishes to thank Alain Miniussi for maintenance and re-factorisation of the code FARGOCA. 
\end{acknowledgements}

%
%
\bibliographystyle{aa} 
\bibliography{references}
\begin{appendix} 
\section{Comparison between NIRVANA and FARGO3D}
\label{sec:Appendix}
Here we present a comparison between the NIRVANA and FARGO3D codes when applied to a standard disc-planet interaction scenario. A run was performed using the disc model and setup described in Sect.~\ref{sec:Methods}. We adopted a model without an accreting layer and with viscous parameter $\alpha=10^{-5}$, as in the No Wind case that we presented in the main paper. A 20~M$_{\oplus}$ planet was inserted in the disc at 5.2~au on a fixed circular orbit and the radial mass flux in the disc was computed in the same way as was done for the mass fluxes presented in the right panel of Fig.~\ref{fig:mdot-Mp300}. Figure~\ref{fig:NIRV+FARGO3D} shows the results obtained by the two codes, where the lines represent the azimuthally and meridionally integrated mass fluxes at each radius, averaged between the times corresponding to 400-480 planet orbits. As expected, the outward and inward propagating spiral waves lead to an angular momentum flux, and an associated mass flux, that is concentrated around the planet where the waves damp. Further from the planet, the mass flux essentially disappears because a wind torque is not applied in this case and the viscosity is very small. Good agreement is obtained between the two codes.
\begin{figure}
\includegraphics[width=\columnwidth]{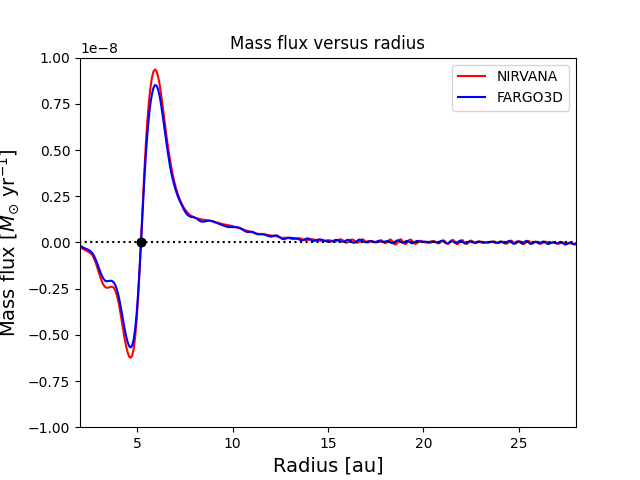}
\caption{Time-averaged mass flux versus radius for runs performed using NIRVANA and FARGO3D.}
\label{fig:NIRV+FARGO3D}
\end{figure}
\end{appendix}
\end{document}